# Stereodynamic Imaging of Bromine Atomic Photofragments Eliminated from 1-Bromo-2-Methylbutane Oriented via Hexapole State Selector

Masaaki Nakamura[1], Hsiu-Pu Chang [1], King-Chuen Lin[1,2,*], Toshio Kasai[1,3], Dock-Chil Che[4], Federico Palazzetti[5], and Vincenzo Aquilanti[5,6]

[1] *Department of Chemistry, National Taiwan University, Taipei 10617, Taiwan*

[2] *Institute of Atomic and Molecular Sciences, Academia Sinica, Taipei 10617, Taiwan*

[3] *Institute of Scientific and industrial Research, Osaka University, Ibaraki, Osaka 567-0047, Japan*

[4] *Department of Chemistry, Graduate School of Science, Osaka University, Toyonaka, Osaka 560-0043, Japan*

[5] *Università di Perugia, Dipartimento di Chimica, Biologia e Biotecnologie, 06123 Perugia, Italy*

[6] *Istituto di Struttura della Materia, Consiglio Nazionale delle Ricerche, 00016 Rome, Italy.*

*To whom correspondence should be addressed at kclin@ntu.edu.tw

# Abstract

Both single-laser and two-laser experiments were conducted to look into the ion-imaging of Br*($^2P_{1/2}$) and Br($^2P_{3/2}$) photo-fragmented from 1-bromo-2-methylbutane in the range 232-240 nm via a detection scheme of (2+1) resonance-enhanced multiphoton ionization. The angular analysis of these photofragment distributions yields the anisotropy parameter $\beta = 1.88 \pm 0.06$ for the Br* excited state which arises from a parallel transition, while $\beta = 0.63 \pm 0.09$ for the Br ground state indicates the contribution from both a perpendicular transition and a non-adiabatic transition. When a hexapole coupled with an orienting field was implemented, the parent molecules are spatially oriented to yield an orientation efficiency $|\langle\cos\theta\rangle|$ of 0.15. Besides the $\chi$ angle between the recoil velocity **v** and the transition dipole moment **μ**, orienting molecules allows for the evaluation of the angle $\alpha$ between **v** and the permanent molecular dipole moment **d**. The angular analysis of Br* photofragment distribution yields $\chi$ to be 11.5° and $\alpha$ in the range from 160° to 180° with weak dependency. In the two-laser experiments, the angular anisotropy of Br photofragment distribution was found to be smaller ($0.38 \pm 0.10$) when the photolysis wavelength was red-shifted to 240 nm, suggesting the increasing contributions from perpendicular transitions.

## I. Introduction

In gas-phase molecular photo-fragmentation, the analysis of vector properties has been made feasible by implementation of ion imaging. The information on vector properties offers a detailed insight into the related molecular stereodynamics such as alignment and orientation, photo-excitation routes, dissociation lifetime, coherence in electronic state, and dissociation mechanisms. Regarding the photofragment scattering angular distribution via one photon absorption, the following equation is usually employed to characterize the anisotropy parameter, $\beta$,

$$I(\theta) = \frac{1}{4\pi}[1 + \beta P_2(\cos\theta)] \tag{1}$$

where $P_2(\cos\theta)$ is the second order Legendre polynomials.[1] $I(\theta)$ is the three-dimensional angular distribution function; since it is cylindrically symmetric around the light polarization axis, it depends on only the angle $\theta$ between the recoil direction and the light polarization axis. When the photodissociation is triggered by a single electronic transition, the anisotropy parameter $\beta$ is related to the angle $\chi$ between the recoil velocity direction and the transition dipole moment,

$$\beta = 2P_2(\cos\chi) \tag{2}$$

This procedure enables extraction of the $\chi$ angle, which is essential to characterize the photodissociation dynamics, particularly for symmetric structures such as those of diatomic and symmetric top molecules. For these molecules, geometrical relationships are established among the transition dipole moment, the molecular symmetry axis and the permanent dipole moment. Under the axial recoiling approximation, the recoil direction is tightly connected to the molecular symmetry axis, and thus all the vector quantities involved can be described readily. However, when a molecule lacks the symmetry axis, for instance, when the molecule contains some functional groups, the vector correlations become more complicated, since the recoil velocity, the transition dipole moment and the permanent dipole moment can be aligned diversely in space.

It has been considered a challenging task to characterize the photodissociation dynamics for symmetry-breaking systems. Recently, Janssen and his co-workers looked into the photodissociation behavior of vibrationally excited OCS to show the non-axial recoiling of CO fragments.[2,3] In these experiments, the permanent dipole moment of parent OCS molecules was spatially oriented by a hexapole state selector and an orientation field. Given the fixed geometry of the molecular orientation, the directional preference of the recoiling fragments may be regulated following polarized laser photolysis, so that the resulting photofragment ion images appear asymmetric. Besides the above-mentioned $\chi$ angle, the spatial constraint imposed by molecular orientation allows for further insight into the geometric relationship between the fragment recoiling direction and the permanent dipole moment. Since the geometry of the permanent dipole is usually static with respect to the molecular frame, the behavior of non-axial recoiling in molecular photolysis can be comprehended with the aid of this angle parameter. Despite stereodynamic complexity for the asymmetric molecular dissociation, Rakitzis et al. established a theoretical treatment for the photofragment distributions eliminated from spatially oriented molecules.[4] The expression is formulated by the practical set of apparatus parameters and three geometric angles in the recoil frame, and is generally applicable to larger asymmetric systems.

Recently, our group reported the first example on photodissociation of the hexapole-oriented asymmetric-top molecule 2-bromobutane.[5–7] Despite the lack of symmetry axis, inspecting the time-of-flight (TOF) spectra of $^{79}$Br and $^{81}$Br fragments allowed for a clear visualization of spatial molecular orientation via a biased hexapole state selector in conjunction with an orientation field. The acquired Br($^2P_{3/2}$) (or Br*($^2P_{1/2}$)) ion images appeared to be up-down asymmetric as a result of the molecular orientation; these results enable extraction of the recoil frame angles. Accordingly, the orientation control technique has been demonstrated as a powerful tool to study the photodissociation dynamics in asymmetric systems.

In this work, we conducted photodissociation of 1-bromo-2-methylbutane (BrCH$_2$CHCH$_3$CH$_2$CH$_3$) oriented via hexapole state selector, a further asymmetric-top molecule with even larger size than 2-bromobutane. Thus far, such a large asymmetric top molecule has never been treated before and whether its structure can be fixed spatially is yet to be known. Despite its larger size with branched alkyl chain, fortunately, we found the molecule shows the feature of appreciable orientation after implementation of hexapole field. As for the measure of the molecular orientation efficiency, the ensemble average of the angle between the molecular permanent dipole and the orienting electric field |<cos$\theta$>| is estimated to be 0.15. Upon irradiation of the oriented molecules with the laser wavelengths selected at either 231.9 or 233.6 nm, the spin-orbit excited state Br$^*$ ($^2$P$_{1/2}$) and the ground state Br ($^2$P$_{3/2}$) are ionized individually via a (2+1) resonance-enhanced multiphoton ionization (REMPI) scheme,[8] followed by sliced-ion imaging acquisition.

The Br and Br* fragments have different symmetric correlation with the excited A band of the parent molecule. The Br$^*$ photofragments almost exclusively result from a parallel transition, whereas the Br photofragments arise from both perpendicular transition and non-adiabatic coupling with the parallel transition.[9] As reported there, the elimination pathway of each bromine atomic fragment can be characterized by the set of recoil frame angles. Therefore, given the Br and Br* photofragment distributions, the photodynamical complexity in the photolysis process can be classified by the aid of theoretical simulations. The hexapole orientation technique implemented in the photodissociation dynamics studies should open up the view to look into the vector properties in the molecule-fixed recoil frame, even for larger asymmetric top molecules.

The analysis of sliced ion images yields the correlation of photofragment recoil velocity **v**,

transition dipole moment **μ**, and permanent dipole moment **d**, which are confined by two polar angles, $\chi$, defined above, and $\alpha$, the angle between **v** and **d**, and one azimuthal angle $\varphi_{\mu d}$. Since 1-bromo-2-methyl butane is a chiral molecule, the three-vector correlation plays a vital role for achieving chiral discrimination; but thus far this is still a challenge with a linearly polarized laser as the photolysis light source. In view of this perspective, a simplified formula is presented to comprehend this feature of manifestation of chirality.

## II. Experimental Methods

The photofragment velocity-mapped imaging apparatus has been described elsewhere.[5] In brief, it consisted of three chambers, housing respectively the beam source, the hexapole, and the detection system. The 1-bromo-2-methylbutane enantiomer in S-form produced a vapor pressure of 25 Torr at room temperature passing through a pulsed valve (General Valve Co.) without carrier gas and then expanded into the source chamber. The valve with a 0.6-mm diameter orifice produces pulses of a 280 μs duration operating synchronously with the laser pulses at 10 Hz. After passing through a 1-mm diameter skimmer and a collimator, the molecular beam entered a chamber equipped with the hexapole state selector in which the molecules were focused with selection of specific rotational levels. The hexapole state selector comprised six stainless steel rods each with 4-mm diameter and 70-cm long, and the inscribed circle radius of the hexapole was 4 mm. Its exit was at a distance of 2 cm from the entrance ion lens. The small distance permitted the adiabatic passage of the state-selected molecules into the orienting field without losing the alignment.

Prior to intersection with the polarized laser beam inside the first-stage region of a four-plate ion lens, the molecules were oriented by an orienting field to have the permanent dipole pointing towards either forward or backward direction along the TOF axis. The extraction plate of the ion lenses played two roles, to function both as an orienting field for molecular orientation and subsequently as the ion extraction field for velocity mapping. If the backward orientation is endorsed, a biased pulse should be implemented inside the ion lens to reverse the direction.[5]

A 308 nm XeCl excimer laser (LPX 200, Lambda Physik) -pumped dye laser (PD 3000, Lambda Physik) operating at 10 Hz was used as the photolysis light source. Its output was frequency-doubled to emit at 231.9 or 233.6 nm for the (2+1) REMPI detection of Br*($^2P_{1/2}$) and Br($^2P_{3/2}$) with the energy ranging from 600 to 1000 μJ/pulse, and then softly focused onto the leading edge of the oriented beam with a 300-mm focal length lens to minimize the cluster formation. The radiation was linearly polarized: the polarization was perpendicular to the flight tube and parallel to the detection surface. Prior to propagation into the vacuum chamber, the angle of the laser polarization was adjusted to rotate at 90º or 45º via a half-wave plate with respect to the TOF axis.

The resulting ions were extracted and accelerated into an about 30-cm long field-free drift tube along the molecular beam direction. The ion-cloud expansion was mapped onto a two-stage microchannel plate (MCP) coupled with a phosphor screen (FM3040, Galileo). The MCP was gated with a pulse width of 500 ns to cover all the bromine isotope variant or within a minimum duration of 25 ns for acquisition of sliced ion images. The ion imaging on the phosphor screen was recorded by a charge coupled device (CCD) camera (200XL4078, Pixelfly). Meanwhile, the photon signal was also acquired by a photomultiplier tube (PMT) and then transferred to a transient digitizer to monitor the signal intensity simultaneously. Each image was accumulated at 80,000 – 100,000 laser shots, and the background was removed by subtracting a reference image collected at off-resonance wavelength under otherwise the same conditions. The fundamental laser wavelength was scanned back and forth over a range of ±0.003 nm to detect all the scattering fragments with slight difference of Doppler shift in the laboratory frame.

In order to remove the possibility of alignment effect, the two-laser experiments were conducted. The additional third harmonic of a Nd:YAG laser (Quanta-Ray Lab-170, Spectra-Physics) pumped dye laser (ScanMate Pro, Lambda Physik) at 10 Hz substituted the photolysis light source emitting at 240.0 nm with ~300 μJ/pulse after frequency-doubling. Additionally, the excimer laser-pumped dye laser system was used to probe (2+1) REMPI Br*($^2P_{1/2}$) or Br($^2P_{3/2}$) at 266.55 nm or 266.61 nm with ~900 μJ/pulse, respectively. The probe laser was linearly polarized in parallel with the flight tube to avoid any anisotropy of photofragment distributions, although it has been confirmed that the photofragment distributions were independent of the probe laser polarization. Both beams were in antiparallel alignment. The probe beam was tightly focused with a 25 cm lens, whereas the photolysis beam was mildly focused with a 50 cm lens to cover the probe laser spot spatially. The photolysis-probe beams were adjusted to have a 15 ns delay.

The efficiency of molecular orientation was evaluated by the TOF measurements, which were recorded directly via the MCP, instead of the PMT to circumvent a long-tailed decaying caused by the phosphor lifetime. The TOF spectra were acquired both with and without the hexapole voltage which was biased up to 5 kV. The difference eliminates the effect of the direct beam, thus ensuring the molecular orientation of 1-bromo-2-methyl butane molecule focused by the hexapole (otherwise use is made f cut off by a beam stop).

### III. Results and Discussions

#### A. Molecular orientation

The electric field generated by hexapole state selector may focus the molecular beam along its axis, aligning and selecting the molecules with specific rotational states. Its intensity depends on the factors including rod length, hexapole radius, and the voltage biased to the hexapole. The Stark interaction between the electric field and the permanent dipole moment of parent molecule leads to non-statistical rotational population of the molecules, which may subsequently be oriented either forward or backward with respect to the orienting field. The obtained focusing curve is characterized as a function of the hexapole voltage biased up to the maximum 5.0 kV, yielding a molecular population intensity to increase gradually with the voltage. Unlike the case of linear-top, symmetric-tops and light nearly symmetric-top molecules, the very dense manifold of the rotational levels for the molecule under study does not allow for the selection of a single rotational state. However, the theoretical fit of the observed focusing curve provides the information on rotational temperature and the rotational states selected. For instance, as reported previously, 2-bromobutane comprises three stable conformers,[10] for all the molecular trajectory was simulated appropriately so that their sum reproduces the experimentally measured focusing curve. Accordingly, the total angular momentum was evaluated up to the maximum value of $J$=22 with the rotational temperature of 30 K.

The 1-bromo-2-methylbutane molecule has nine conformers[11]: for all the structures calculated at HF/6-311+G (d, p) level of theory for its structure with the corresponding energy and the relative population at 300 K, as listed in Table 1. Among them, seven conformers have significant contributions to the beam intensity as a function of the hexapole voltage. For simplicity, we conducted the simulation of molecular trajectories[12,13] only for the most stable G-G+ conformer, because of the heavy and complicated computational procedures. The employed simulation parameters are the same as for the 2-bromobutane.[10] As shown in Fig.1a, the plot showing the focusing behavior for 1-bromo-2-methylbutane was obtained by integrating the TOF spectra of all the Br isotopic variants following photodissociation of the oriented molecule. Both curves increase at around 1 kV hexapole voltage and agree well up to 3 kV. The simulation underestimates the beam intensity over 3 kV; other conformers and higher rotational states may need to be included for better reproduction of the experimental results.

Fig.1b shows the TOF spectra of Br$^*$ photofragments with a 5.0 kV hexapole voltage. The spectrum with hexapole voltage off corresponding to the direct beam contribution was subtracted. Note that the laser polarization was set in parallel with the TOF axis, and consequently with the orientation field. The Br* photofragments were produced almost solely in a parallel transition ($\beta$ = 1.88 ± 0.02), and thus distributed forward and backward with respect to the detector. The populations ratio of these two components along the TOF axis may be used to evaluate the orientation efficiency of the parent molecule upon photodissociation. Because bromine has two stable isotopes with almost equal natural abundances ($^{79}$Br:$^{81}$Br = 50.7:49.3), Fig.1b (gray solid line) reveals two sets of double peaks in the spectrum, which are predicted consistently by the Monte-Carlo simulations (the colored dashed lines), showing enhancement of the peaks of each isotope. This observation implies that the Br site of the molecule is oriented toward the detector.

The orientation of the permanent dipole moment is cylindrically symmetric along the orientation field and can be described as a linear combination of Legendre polynomials,[14]

$$O_d(\cos\theta_{dO}) = \sum_{k=0}^{2J} c_k P_k(\cos\theta_{dO}) \qquad (3)$$

$\theta_{dO}$ denotes the angle between the permanent dipole moment and the orientation electric field. The terms up to $k = 2$ are considered here. $c_0$ is 0.5 because of the normalization condition. Since $P_1(\cos\theta)$ is an odd function and $P_2(\cos\theta)$ is an even function, $c_1$ factor represents the orientation while $c_2$ indicates the alignment of molecules. The values $c_1$ = -0.23 and $c_2$ = 0.0 were obtained as a result of fitting the experimental spectrum. The non-zero $c_1$ represents the biased spatial orientation of 1-bromo-2-methylbutane molecules and the sign of $c_1$ expresses the polarity of molecular orientation with respect to the orientation electric field. The conventional representation of molecular orientation <$\cos\theta_{dO}$> is – 0.15 derived as (2/3) $c_1$ using Eq.3.

**B. Vectorial imaging and characterization of the stereodynamic photo-process**

For the photodissociation at 232-233 nm in one-color experiments, Fig.2 shows the sliced images of $^{79}$Br fragments acquired in each state from non-oriented molecules after background subtraction. The image analysis yields the speed and the angular distributions of photofragments, and subsequently the anisotropy parameters following Eq. 1. The Br$^*$ speed distribution peaks at 920 m/s with a standard deviation $\sigma_v$ = 75.0 m/s, while the anisotropy parameter $\beta$ at the peak speed is 1.88 ± 0.06. The obtained $\beta$ is very close to 2, suggesting that almost all the Br$^*$ photofragments should be produced in a parallel transition. In contrast, Br photofragment peaks at 944 m/s with $\sigma_v$ = 101 m/s in the speed distribution, and the $\beta$ is obtained to be 0.63 ± 0.09: this implies the contribution of photodissociation channels from both perpendicular and parallel transitions.

The UV light excitation at ~230 nm involves three dissociative upper states: $^1Q_1$, $^3Q_0$ and $^3Q_1$ in ascending order according to Mulliken.[15] (Fig.3) The excitation only to the $^3Q_0$ surface has a parallel character in direct correlation to the Br$^*$ emission, which is consistent with the observation of sliced imaging experiment. Given the $\beta$ value in photolysis at 231.9 nm, the Br$^*$ photofragment is thus considered to arise from the transition to the $^3Q_0$ surface exclusively. In contrast, the Br photofragment obtained at 233.6 nm yields an intermediate $\beta$, which results from the perpendicular transition to $^3Q_1$ in combination with a non-adiabatic transition from $^3Q_0$ to $^1Q_1$ surface. This path carries a parallel character to result in a positive $\beta$ of Br photofragment, because of the large absorption cross section to $^3Q_0$ surface.[8,9,16–18] The Br speed distribution at 233.6 nm is broader than that of Br$^*$ at 231.9 nm (see similar effect for structural isomers[19,20] and for smaller bromo-alkanes[5,18,21]). Such results are indicative for the nature of dissociative excited states of C-Br bond, thus suggesting that the Br atoms are fragmented slowly on the potential energy surface.

The following equation describes the energy conservation in the photodissociation process,

$$hv + E_{int}^{RBr} - D_0 = E_t + E_{SO}^{Br} + E_{int}^{R} \qquad (4)$$

where $h\nu$ is the photon energy, $E_{int}^{RBr}$ is the internal energy of the parent molecule, $D_0$ is the C-Br bond dissociation energy, $E_t$ is the total translational energy of the photofragments, $E_{SO}^{Br}$ is the spin-orbit splitting energy of Br atom, and $E_{int}^{R}$ is the energy distributed into the internal degrees of freedom of alkyl radical ($C_5H_{11}$). Because of the cooling in adiabatic expansion of parent molecule, $E_{int}^{RBr}$ can be regarded to be negligible. The spin-orbit splitting energy $E_{SO}^{Br}$ is 0.0 kJ/mol for Br and 44.0 kJ/mol for Br*. There is no available data of the C-Br bond dissociation energy $D_0$ of 1-bromo-2-methylbutane (Br-CH$_2$CH(CH$_3$)CH$_2$CH$_3$); however, there are comprehensive data of bromoalkanes with similar structures including 1-bromobutane (Br-CH$_2$CH$_2$CH$_2$CH$_3$, $D_0$ = 296.6±4.2 kJ/mol), 1-bromo-2-methylpropane (Br-CH$_2$CH(CH$_3$)$_2$, $D_0$ = 295.0 kJ/mol), 1-bromopentane (Br-CH$_2$CH$_2$CH$_2$CH$_2$CH$_3$, $D_0$ = 295.0±4.2 kJ/mol), and 1-bromo-2,2-dimethylpropane (Br-CH$_2$C(CH$_3$)$_3$ $D_0$ = 298.2 kJ/mol).[22] The C-Br bond dissociation changes only a few kJ/mol, independent of the size and branching of the saturated carbon chain. While considering the similarity with 1-bromo-2-methylpropane, $D_0$ for 1-bromo-2-methylbutane may well be adopted to be 295.0 kJ/mol. The available energy of photofragments $E_{avl}$ is readily expressed as,

$$E_{avl} = h\nu + E_{int}^{RBr} - D_0 - E_{SO}^{Br} = E_t + E_{int}^{R} \qquad (5)$$

The average of translational energy distribution of Br atom photofragment $\langle E_t^{Br} \rangle$ is obtained from the speed distribution in Fig. 2 and the averaged total translational energy of photofragments $\langle E_t \rangle$ can be derived from the both energy and momentum conservations,

$$\langle E_t \rangle = \frac{m_{RBr}}{m_R} \langle E_t^{Br} \rangle \qquad (6)$$

where $m_X$ represents the mass of parent molecule and alkyl radical. Table 2 shows the $E_{avl}$, $\langle E_t \rangle$, and $\langle E_{int}^{R} \rangle$ of each photofragment. Additionally, the fractions of the energy partitioning into the internal energy of the alkyl radical fragment $f_{int}$ are calculated by the flowing equation and the results are also listed in Table 2.

$$f_{int} = \frac{E_{int}^R}{E_{avl}} \tag{7}$$

In comparison with the excited state Br*, the ground state Br photofragment has larger energy partitioning into the internal degrees of freedom of the co-product alkyl fragment. The fact shows the same trend as the imaging profiles acquired for both states.

The speed distributions in Fig.2 show an apparent peak that is accompanied by a slower component. These two components, the blue and grey lines, were deconvoluted by fitting the speed distribution with Gaussian functions; their summed curve is in red. The strongest peak corresponds to the $^{79}$Br photofragments by judging from the time-of-flight, while the slower component is ascribed to the contribution of the other $^{81}$Br isotope. Because the $^{79}$Br and $^{81}$Br peaks are not perfectly separated (Fig.1b), the leading edge of $^{81}$Br Newton sphere might be resided in the center slice of $^{79}$Br. It should be noted that the $^{81}$Br contribution lies in the central region of the image, but independent of determination of the peak position in the $^{79}$Br speed distribution.

In non-oriented experiments, the $\chi$ angle used to characterize the geometry of the recoiling direction and the transition dipole moment is achieved by analyzing the fragment images (Fig.4). Additionally, when molecules are oriented, the geometry of the permanent dipole moment also affects the photofragment distribution, enabling to estimate the angle $\alpha$ from experimental images. For the symmetric top molecules, **d** is simply along the molecular symmetric axis. However, for the asymmetric top molecules like 1-bromo-2-methylbutane, for which the permanent dipole moment **d** is not necessarily parallel to the recoil velocity **v**, the angle $\alpha$ appears neither 0° nor 180° even in the equilibrium structure. The permanent dipole moment is closely associated with the molecular skeleton, almost independent of the photolysis wavelength: thus, it can be used as a good reference to clarify the photodissociation dynamics under symmetry breaking conditions.

Since the orientation field is parallel to the TOF axis, the effect of molecular orientation on the photofragment imaging disappears with the laser polarization aligned at 0° (horizontal) or 90° (vertical). For this reason, the laser polarization is titled to 45° with respect to the TOF axis and the sliced images of Br* photofragments were subsequently acquired at either 0 or 5 kV hexapole voltage, as shown in Fig.5. Images obtained at 0 kV permits to subtract the background under the off-resnance condition, while images at 5 kV permit to subtract the background at 0 kV. The molecules at 0 kV hexapole voltage are in spatially isotropic distribution; however these molecules become oriented at 5 kV (Fig.1b). Including the tilted polarization, the photofragment angular distribution from oriented parent molecules is formulated as an expansion of associated Legendre polynomials,[7]

$$I(\theta) = 1 + \beta_1^0 P_1^0(\cos\theta) + \beta_2^0 P_2^0(\cos\theta) + \beta_1^1 P_1^1(\cos\theta) + \beta_2^1 P_2^1(\cos\theta) \quad (8)$$

Where the $P_l^m$ are the associated Legendre polynomials and the corresponding coefficients $\beta_l^m$ are determined by $b_l^m/b_0^0$. $P_2$ is the second order Legendre polynomial, namely $P_2 = P_2^0$.

$$b_0^0 = (1 - c_2 P_2(\cos\alpha))\left(1 - \frac{1}{2}P_2(\cos\chi)\right) + \frac{3}{16}c_2 \sin^2\alpha \sin^2\chi \cos 2\varphi_{\mu d} \quad (9)$$

$$b_1^0 = 3c_1 \sin\alpha \sin\chi \cos\chi \cos\varphi_{\mu d} \quad (10)$$

$$b_2^0 = (1 - c_2 P_2(\cos\alpha))P_2(\cos\chi) + \frac{3}{8}c_2 \sin^2\alpha \sin^2\chi \cos 2\varphi_{\mu d} \quad (11)$$

$$b_1^1 = -\frac{9}{8}c_2 \sin^2\alpha \sin^2\chi \sin 2\varphi_{\mu d} \quad (12)$$

$$b_2^1 = -c_1 \sin\alpha \sin\chi \cos\chi \sin\varphi_{\mu d} \quad (13)$$

As discussed above, the Br* photofragments are expected to arise solely from the $^3Q_0$ surface. Thus, the photofragment angular distribution is characterized by only a set of ($\alpha$, $\chi$, $\varphi_{\mu d}$).

$$I_{Br^*} = I_\parallel(\alpha_\parallel, \chi_\parallel, \varphi_{\mu d\parallel}) \quad (14)$$

The angle $\chi$ can be determined by the anisotropy parameter $\beta$ as described in Eq. 2, from $\beta = 1.88\,\chi_{//}$ can be determined to be either 11.5° or 168.5°. The two angles are complementary (their sum is 180°) and cannot be distinguished from each other, since the electric field vector of the laser beam is oscillating. The azimuthal angle $\varphi_{\mu d //}$ is also an essential parameter to describe the recoiling frame. However, the $\varphi_{\mu d}$ dependency is limited from geometric consideration if $\chi \sim 180°$; if the transition dipole moment **μ** is located close to the recoil velocity **v**, the geometry confined by the **v-d-μ** vectors becomes essentially similar for any $\varphi_{\mu d}$. Therefore, we assigned $\varphi_{\mu d}$ to be 0°. Given the angle $\chi$ and $\varphi_{\mu d}$, the angle $\alpha$ is then determined from the experimental finding. Accordingly, the $\alpha$ angle derived from the photofragment distributions is found to lie in between 160° and 180°, this range yielding reasonable comparison between the experimental and calculated results. It is not easy to pin down a single value for the best fit because of the poor dependency in the angular region. The green band overlapped on the angular distribution of Br$^*$ in Fig. 5 shows the range of simulated angular distributions with changing $\alpha$ from 160° to 180°. The dependence on the $\alpha$ angle may be enjanced by increasing the degree of orientation, namely, by increasing $c_1$.

As confirmed theoretically, different forms of enantiomers are predicted to give an angular shift on the photofragment distribution depending on the geometry of **v-d-μ** vectors.[2>] This phenomenon is mainly caused by the terms $\beta_1^1$ and/or $\beta_2^1$ in Eq. 8, which are the coefficients of odd functions.[7] These coefficients are proportional to either $c_1$ or $c_2$ parameter, and thus they can be non-zero only if the parent molecules are oriented or aligned. For the case of 1-bromo-2-methylbutane molecule, there was no significant shift on angular distributions between oriented and non-oriented molecules (Fig.5) This observation weakly supports the assumption of $\varphi_{\mu d} \sim 0°$. The conditions for $\alpha$ and $\chi$ are also critical, especially the $\chi$ imposes strong constraints. However, under the axial recoil approximation, $\chi$ is typically obtained close to the value of $\sim 0°$ or $\sim 90°$, which is not favorable to visualize the angular shift. Despite a slower process for the Br fragmentation, the angular distributions show irrelevant difference between the conditions with and without molecular orientation. Regarding violation of the axial recoil approximation, an example is reported previously for the OCS molecule in which the vibrational states were excited.[2,3] In addition, it is arguable that selection of a predissociative upper state prolongs the dissociation lifetime.

Fig.6 shows the Br and Br* ion images under generally the same condition, but with the gate pulse width of the detector increased to 800 ns to cover the total projection images. The obtained image of Br* (Fig.6a) shows stronger effect of molecular orientation featuring as the up-down asymmetry. The up-side stronger fashion correlates the recoiling vector **v** and the permanent dipole moment **d**, implying that $\cos(\alpha) < 0$ in weak field limit orientation, similar to the work by Rakitzis et al.[2,3] It is assumed that the permanent dipole points from the negative charge side to the positive side. The angle $\alpha$ is found to be similar to the angle between the C-Br bond and the permanent dipole moment under axial recoiling approximation. We calculated the equilibrium geometries of nine rotamers of 1-bromo-2-methylbutane based on MP2/6-311 G + (d,p) level and tabulated the results in Table.1. Although there are slight differences, similar values approximately appear from 168° to 171°. Fig.6b shows the result of a Monte-Carlo simulation with $(\alpha, \chi) = (169.4°, 11.5°)$; we chose $\alpha = 169.4°$ since the G-G+ conformer is the most stable. The simulated image agrees well with the experimental result in Fig.6a.

The measured Br photofragment distribution is shown in Fig.6c. The image becomes more isotropic because of a smaller $\beta$, but still exhibiting a slight asymmetric feature. Since the intermediate $\beta$ implies the existence of both parallel and perpendicular characters, the photofragment distribution is modeled as the sum of these two contributions.

$$I_{Br} = c_\parallel I_\parallel(\alpha_\parallel, \chi_\parallel) + c_\perp I_\perp(\alpha_\perp, \chi_\perp) \tag{15}$$

where $c_\parallel$ and $c_\perp$ are the weighting factors for each contribution. Thus, the above equation may be further expressed as,

$$c_\parallel + c_\perp = 1 \tag{16}$$

$$\beta_{Br} = c_\parallel \beta_\parallel + c_\perp \beta_\perp \tag{17}$$

Now, $\beta_{Br}$ is 0.63 and $\beta_\parallel \sim \beta_{Br^*} = 1.88$. $I_\parallel$ should be similar to $I_{Br^*}$, because the photolysis wavelength is close, and only one upper surface is related. Using Eq. 2, $c_\perp$ is expressed as a function of $\chi_\perp$,

$$c_\perp = \frac{\beta_\parallel - \beta_{Br}}{\beta_\parallel - 2P_2(\cos \chi_\perp)} \tag{18}$$

The angle $\alpha$ depends only on its molecular structure, but independent of the transitions if the dissociation process is fast enough. Then the angle $\alpha_\perp$ should be the same as $\alpha_\parallel$ obtained in the Br$^*$ case. Further, we carried out the Monte-Carlo simulation obtaining the best fit of the Br fragment image with $\chi_\perp = 90°$, Fig.6d.

### C. Two-laser experiments

To overcome the limitations of single-laser experiments, the two-laser experiments have been carried out additionally. The photolysis wavelength was red-shifted to 240.0 nm, and then the photofragment Br$^*$ and Br atoms were ionized by a REMPI scheme at 266.51 and 266.66 nm, respectively.

The two-laser setup offers the following merits. First, it brings about flexibility to select excitation wavelength, without any restriction by the probe wavelength used in a REMPI scheme, like the case in a single-laser experiments. The tunable photolysis wavelength allows for controlling the available energy for the bond dissociation and the subsequent fragment energy distributions from each dissociation path. Second, it is possible to decrease the photolysis laser power, since high photon density demanded for (2+1) REMPI is no longer required. Thus, the possibility of laser light alignment effect is reduced. Third, the probe wavelength is allowed to select freely a stronger REMPI line for improving the signal-to-noise ratio of acquired images. Note that the absorption cross section for the photolysis laser beam is desirable to be larger than that of probe beam. In the two-laser experiments, the probe wavelength is tuned to ~266 nm, at which the absorption cross section of the parent molecule is so small that the REMPI signal could not be observed only with the probe laser.

Regarding the photodissociation at 240 nm, $\beta$ was obtained to be $1.81 \pm 0.09$ for Br$^*$ probed at 266.55 nm and $0.38 \pm 0.10$ for Br at 266.61 nm. The change of photolysis wavelength should cause deviation in the angular distributions of photofragments. As expected, the photolysis at 240 nm may lead to the increased contribution of the perpendicular character to the Br fragment, in comparison to the result at 233.6 nm.

As shown in Fig.7, the sliced images of photofragments from oriented molecules were measured under otherwise the same condition as in the one-laser experiments. The Br$^*$ result appears similar to that in the single laser experiments (Fig.5), whereas the angular distribution of Br fragments is more isotropic than that in Fig.5. The difference between the angular distributions obtained with and without hexapole involvement is irrelevant with that at 240 nm photolysis wavelength; thus, the recoil frame geometry is expected to be similar to the result at 233.6 nm. However, it is worth noting that the signal-to-noise ratio of the Br photofragment is improved.

Although the gate pulse to the MCP detector is aimed at the center slice of $^{79}$Br photofragment distribution, the other isotope $^{81}$Br photofragments contribute to the central region of the images. After careful comparison of the sliced imaging results and the TOF measurements, we concluded that the sliced images in Fig. 7 are the observation of the $^{79}$Br photofragments, while partial $^{81}$Br photofragments contribute to the central images. Fig. 2 also shows similar results. The center signal is enhanced much more than those in Fig. 2, because of the tilted laser polarization. The leading edge of the $^{81}$Br Newton sphere is extended into the center slice of $^{79}$Br photofragments and contributes to the central region of images. When the linear polarization is tilted at 45º with respect to the TOF axis, the polar region of the Newton sphere comes closer to the leading edge, which has stronger photofragment distribution than the equatorial region as suggested by positive anisotropy parameter *β*. Furthermore, the photofragment distribution at the leading edge increases with approaching to the isotropic pattern at *β* equal to 0. That is why the $^{81}$Br Newton sphere contributing to the central images becomes more significant for the Br photofragments. The bimodal images cannot be associated with the direct and the non-adiabatic transition pathways which yield very close kinetic energy releases.

## IV. Conclusion

The photodissociation reaction of 1-bromo-2-methylbutane around 232-240 nm has been investigated using velocity-mapped imaging setups in both single-laser and two-laser experiments. The hexapole state selector was installed to control the spatial orientation of the skimmed molecular beam prior to photodissociation and the subsequent ion-imaging detection of Br and Br* fragments via a (2+1) REMPI scheme. In spite of the heavy molecular mass and the dense manifold of rotational states, the molecular beam could be appropriately focused by Stark interaction through the hexapole and then oriented by the orienting field. The degree of molecular orientation was evaluated to be $|<\cos\theta>| = 0.15$ in the TOF measurements. The analysis of photofragment ion images yielded the anisotropy parameter *β*, and the subsequent *χ* angle between the recoil velocity and the transition dipole moment. The *β* was obtained to be $1.88 \pm 0.02$ for Br* and $0.63 \pm 0.09$ for Br photofragments. The Br* photofragment is expected to arise solely from parallel transition, while the Br fragments implies the mixed contributions from the perpendicular transitions in conjunction with non-adiabatic transitions.

Furthermore, additionally, the molecular orientation of parent molecules allows for evaluation of the angle $\alpha$ between the recoil velocity and the permanent dipole moment. From the angular analysis of Br$^*$ photofragment distribution, $\chi$ was determined to be 11.5° and $\alpha$ was estimated in the range from 160° to 180° with weak dependency. Using $\alpha$ = 169.4° evaluated by DFT calculation for the most stable conformer under the axial recoiling approximation, the Monte-Carlo simulation reaches the good agreement in the total projection images. The consistency suggests that both the transition dipole moment and the permanent dipole moment should be located close to the axis of the recoil velocity upon photodissociation.

The two-laser experiments were additionally carried out to take advantage of different wavelengths used in the stage of photolysis and ionization. The resulting fragment distributions rule out the possibility of alignment interference which might occur in a single-laser experiment. The setup offers the probability for the photolysis wavelength to cover the A band and produces a better signal-to-noise ratio by selecting detection wavelength. The angular anisotropy of Br photofragment distribution was found to be smaller when the photolysis wavelength was red-shifted to 240 nm, suggesting the increasing contributions from perpendicular transitions. The two-laser experiments implements about the flexibility for the photolysis wavelength and thus enable one to select another dissociative excitation channel beyond the A-band, perspectively permitting chiral recognition.

## V. Acknowledgements

This work is supported by the Ministry of Science and Technology of Taiwan, Republic of China under Contract No. NSC 102-2113-M-002009-MY3. T. K. thanks National Taiwan University for providing him a visiting professorship to carry out this work. The Japanese Ministry of Education,

Science, and Culture is gratefully acknowledged for a Grant Aid for Scientific Research (No. 17KT0008) in support of this work. F. P. and V. A. acknowledge the Italian Ministry for Education, University and Research, MIUR, for financial supporting: SIR 2014 "Scientific Independence for young Researchers" (RBSI14U3VF).

**Table captions**

**Table 1.** *α* angles and relative populations of nine stable conformers of 1-bromo-2-methylbutane at room temperature, calculated at the HF/6-311+G (d, p) level.

**Table2.** Energy partitioning components in the photodissociation of 1-bromo-2-methylbutane.

**Tables**

Table.1

| conformer | Angle [degree] | Relative population at 300 K |
|---|---|---|
| *TT* | 170.4 | 0.409 |
| *TG+* | 169.3 | 0.304 |
| *TG-* | 169.3 | 0.430 |
| *G+T* | 170.7 | 0.355 |
| *G+G+* | 168.8 | 0.004 |
| *G+G-* | 170.4 | 0.125 |
| *G-T* | 170.6 | 0.012 |
| *G-G+* | 169.4 | 1 |
| *G-G-* | 168.9 | 0.267 |

Table.2

| Fragment | $E_{avl}$ [kJ/mol] | $\langle E_t \rangle$ [kJ/mol] | $\langle E_{int}^R \rangle$ [kJ/mol] | $f_{int}$ |
|---|---|---|---|---|
| Br | 217.1 | 75.8 | 141.3 | 0.65 |
| Br* | 176.9 | 71.2 | 105.7 | 0.60 |

**Figure captions**

**Fig. 1.** (a) The focusing curve of 1-bromo-2-methylbutane in which Br* photofragments along the time-of-flight axis were detected following 231.9 nm photolysis. Three conformers, G-G+, TG-, and TT, were taken into account. The rectangular plot shows the experimental results with a 95% confidence interval, while the grey line is the result of the corresponding simulations (See ref. 13). Good agreement is observed in the low voltage region up to about 3 kV. (b) The time-of-flight spectrum of Br* photofragments at 231.9 nm photolysis. The gray line shows the experimental spectrum obtained by the difference of the spectra with 5 kV and with 0 kV hexapole voltages. The colored lines are the Monte-Carlo simulations with orientation coefficients $c_1 = 0.35$ and $c_2 = 0.0$. The red and blue dashed lines correspond to $^{79}$Br and $^{81}$Br, respectively. The green line shows the sum of spectra of those isotopes and agrees well with the experimental result.

**Fig. 2.** The sliced imaging results of $^{79}$Br* and $^{79}$Br fragments from non-oriented molecules and the corresponding speed and angular distributions. The Br fragment presents relatively larger speed distribution, which implies the slower motion on the potential energy surface of dissociative excited states. The weak signal at the center may be caused by partially intruded $^{81}$Br fragments. The Br* fragment distribution leads to an anisotropy parameter near the parallel limit ($\beta = 1.88$), while Br results from a mixing of perpendicular direct dissociations and the non-adiabatic transition from $^3Q_0$ to $^1Q_1$ surface.

**Fig. 3.** Schematic plot of energy surfaces involved in A-band excitation of C-Br bond. Only the transition to $^3Q_0$ surface has a parallel character while the other two have a perpendicular character. The $^3Q_0$ surfaces is correlated with the Br* photofragments by a parallel transition, while the $^1Q_1$

and $^3Q_1$ surfaces are correlated with Br by a perpendicular transition. The $^3Q_0$ surface also contributes to Br fragment via non-adiabatic transition to the $^1Q_1$ surface. This non-adiabatic contribution results in the positive anisotropy parameter for the Br photofragment distribution, since the transition to the $^3Q_0$ surface has a larger absorption cross section. The grey arrows represent schematic energy diagram of photolysis and probe lasers in one- and two-laser experiments.

**Fig. 4.** The scheme of the recoil frame containing three vectors: the recoil velocity **v**, the permanent dipole moment **d**, and the transition dipole moment **μ**. In the *xyz* coordinate, the *z* axis is defined along the **v** vector and *zx* plane contains the **v** and **d** vectors. The geometry of those vectors are confined by three angles ($\alpha$, $\chi$, $\varphi_{\mu d}$). $\alpha$ is the angle between **v** and **d**, and $\chi$ is the angle between **v** and **μ**, while $\varphi_{\mu d}$ is the azimuthal angle of **μ** vector on *xy* plane. Under the axial recoil approximation, the angle $\alpha$ is determined by the parent molecular structure and the angle $\chi$ depending on the excited transition.

**Fig. 5.** Sliced imaging results of oriented molecules with the laser polarization tilted at 45º with respect to the TOF axis in the single-laser experiments at 233.6 nm. Left panel: The obtained images with 0 kV and 5 kV hexapole voltage. The background image is subtracted from 0 kV image and the raw 0 kV image is subtracted from 5 kV image. Right panel: The angular distributions of the images. Both the angular distributions of Br$^*$ and Br fragments present the features similar to the two hexapole voltages, respectively. The result implies that the recoil frame of the photodissociation at this wavelength is close to the symmetric geometry despite the long and branched carbon chain. The angle $\alpha$ of Br$^*$ emission is confined in the range from 160° to 180° with the fitting using Eq. 8. The green band in the upper right panel shows the calculated angular distributions with $\alpha$ varied from 160° to 180°.

**Fig. 6.** The total projection images of oriented molecules with the tilted laser polarization. (a) The experimental image of Br* photofragments featuring the up-down asymmetry caused by parent molecular orientation. (b) The simulated image of Br* fragments with the set of recoil frame angles $(\alpha, \chi) = (169.4°, 11.5°)$. (c) and (d) The experimental and simulated images of Br fragment, respectively. The Monte-Carlo simulation have taken into account the mixing of parallel and perpendicular transitions.

**Fig. 7.** The sliced imaging results of oriented molecules with the tilted laser polarization in the two-laser experiments, in which the conditions and analysis are in common, Fig. 5. Left panel: The experimental findings of Br and Br* images. The photolysis wavelength is at 240 nm, different from 232-233 nm used in the single laser experiment. Their results appear similar, but the signal-to-noise ratio is considerably improved in the Br fragment detection. Right panel: The angular shift in either Br or Br* photofragment angular distributions is irrelevant between the oriented molecules via a hexapole biased at individual 5 kV and 0 kV.

**Figures**

Fig.1

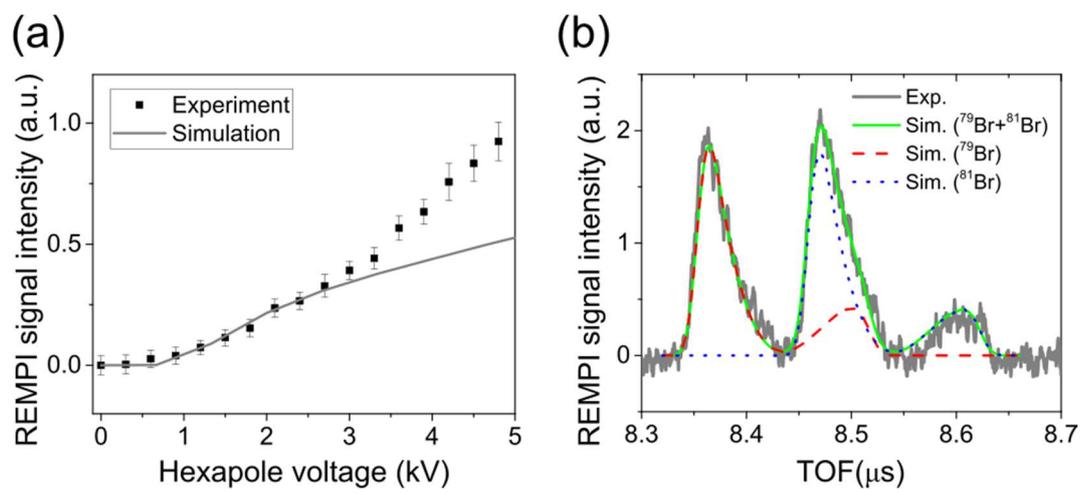

Fig.2

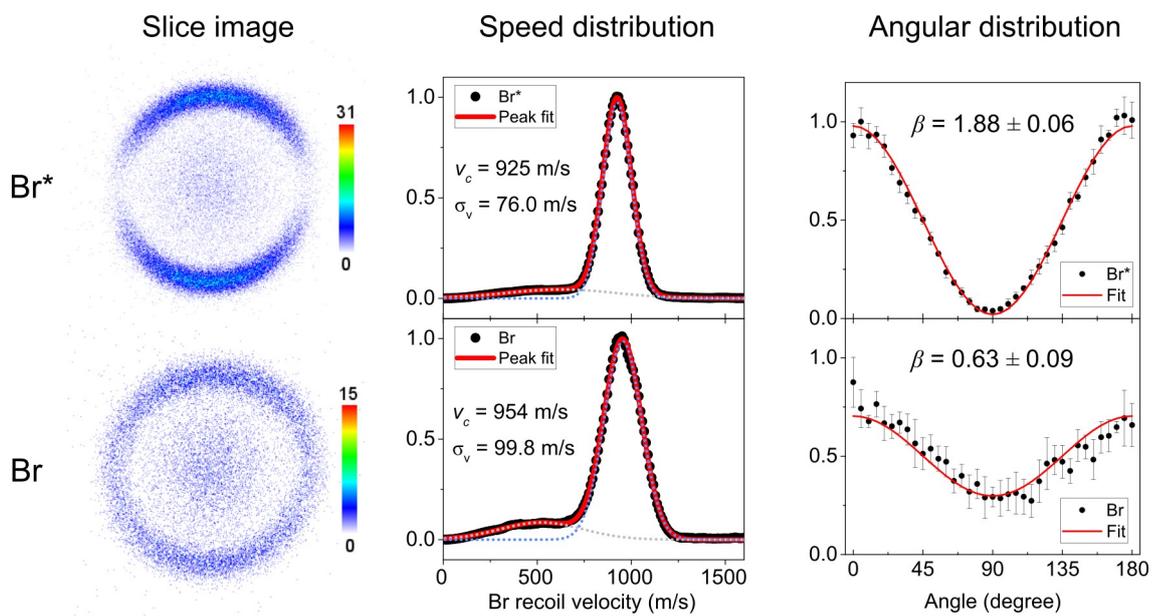

Fig.3

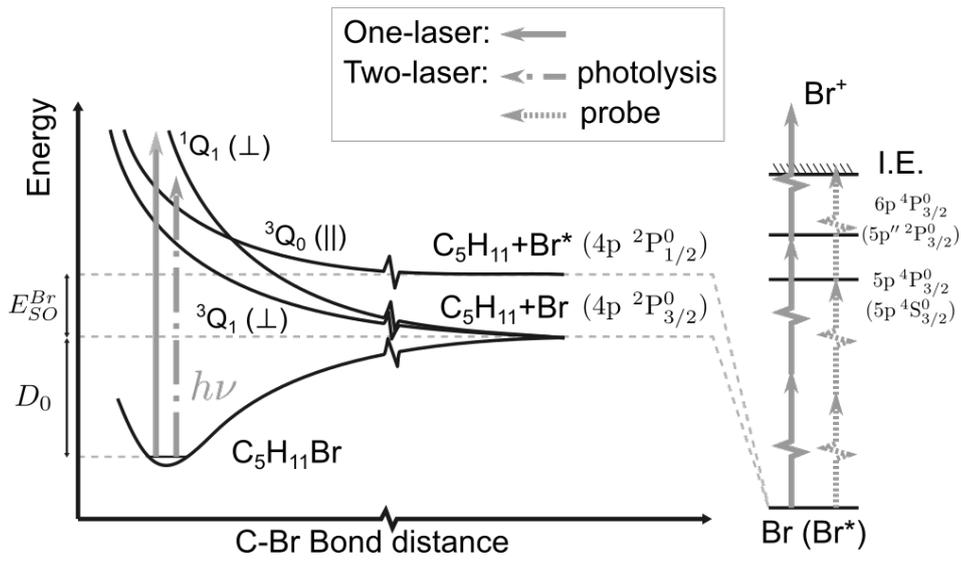

Fig.4

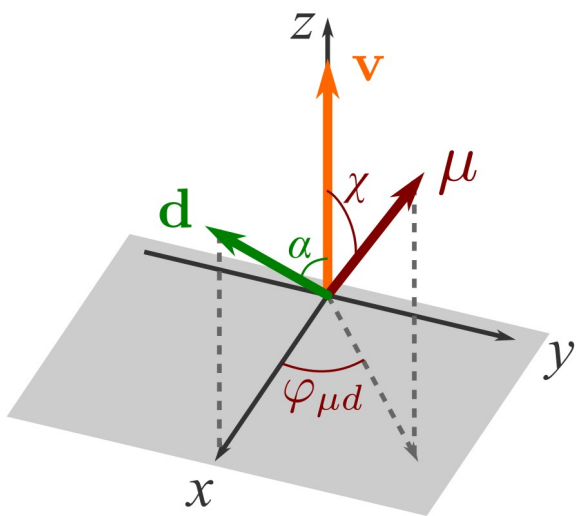

Fig.5

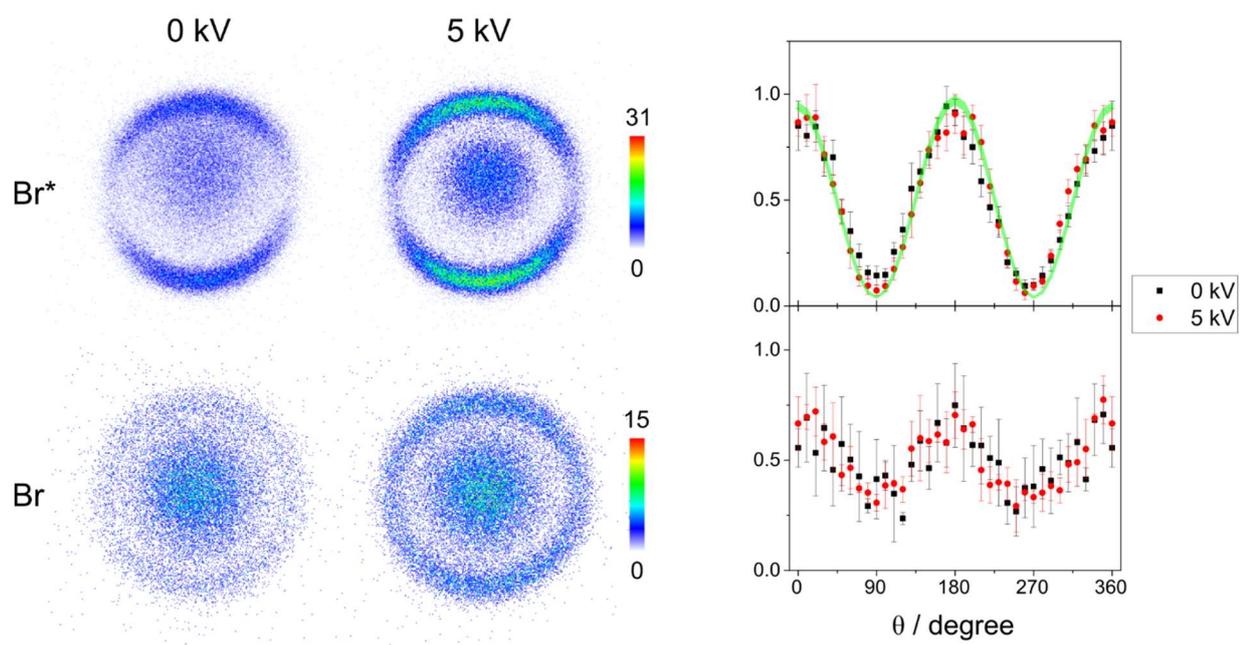

Fig.6

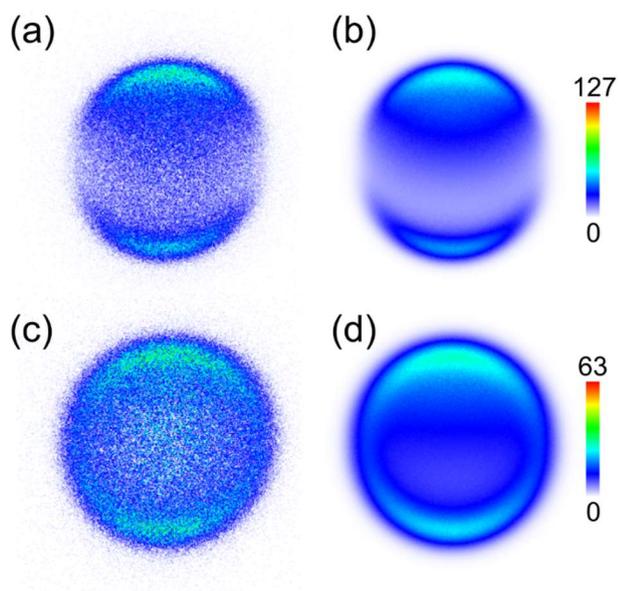

Fig.7

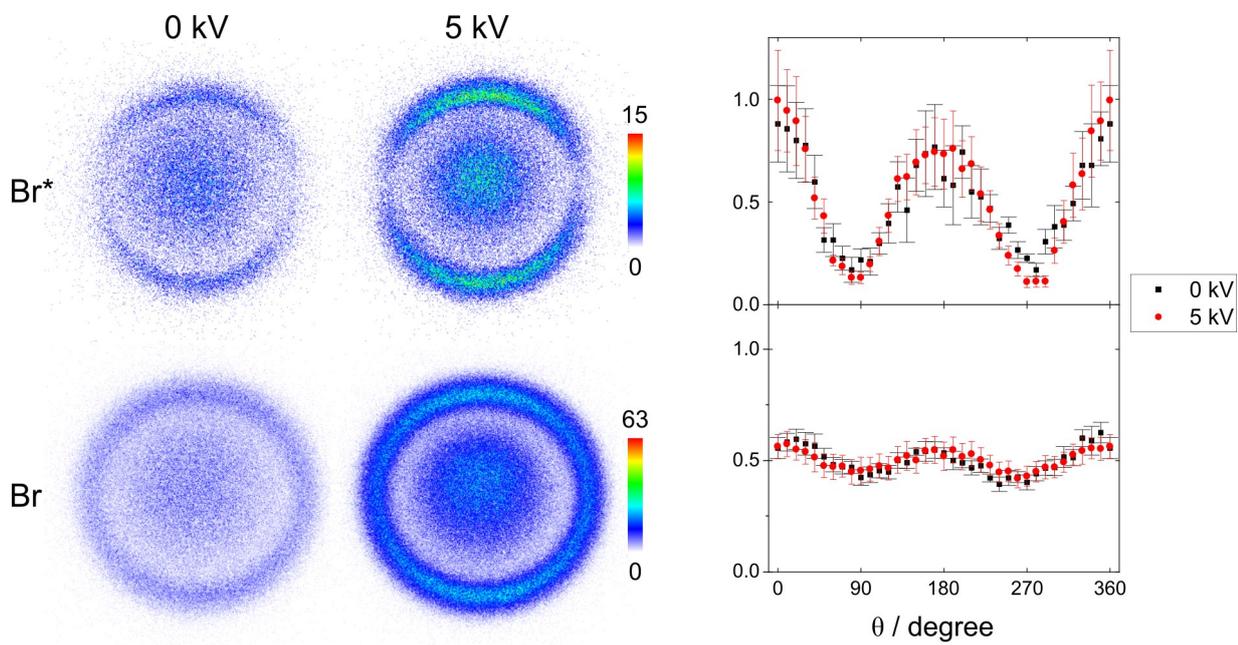

**Table of Contents Graphic**

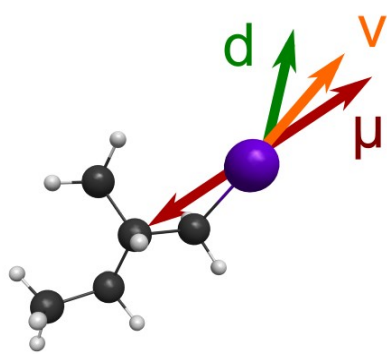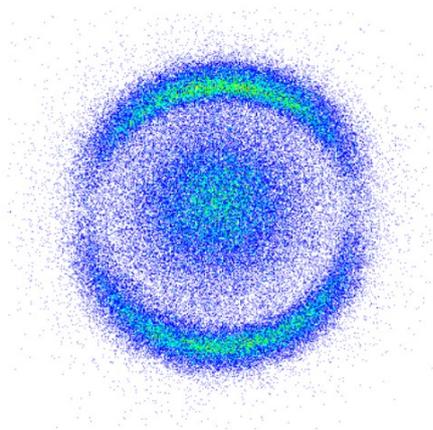